\documentclass[conference]{IEEEtran}
\IEEEoverridecommandlockouts
\usepackage{cite}
\usepackage{amsmath,amssymb,amsfonts}
\usepackage{graphicx}
\usepackage{textcomp}
\usepackage{xcolor}
\usepackage{flushend}
\usepackage{balance}
\usepackage{epsfig}
\usepackage{epstopdf}

\usepackage[linesnumbered,ruled,vlined]{algorithm2e}
\usepackage{float}
\usepackage{subfig}
\usepackage{times}
\usepackage{caption}
\DeclareMathAlphabet{\mathbbold}{U}{bbold}{m}{n}
\usepackage[top=0.71in, bottom=1in, left=0.625in, right=0.625in]{geometry}
\def\BibTeX{{\rm B\kern-.05em{\sc i\kern-.025em b}\kern-.08em
    T\kern-.1667em\lower.7ex\hbox{E}\kern-.125emX}}

\begin{document}

\title{Markov Modeling for Licensed and Unlicensed Band Allocation in Underlay and Overlay D2D\\
\thanks{This work was supported in part by the Academia Sinica (AS) under Grant 235g Postdoctoral Scholar Program, in part by the National Science and Technology Council (NSTC) of Taiwan under Grant 112-2218-E-110-004, 112-2218-E-110-003, and in part by the U.S. National Science Foundation (NSF) under Grant CNS-2030215.}
}
\author{\IEEEauthorblockN{Po-Heng Chou$^{1}$, Yen-Ting Liu$^{2}$, Wei-Chang Chen$^{3}$, and Walid Saad$^{4,5}$}
\IEEEauthorblockA{$^{1}$Research Center for Information Technology Innovation (CITI), Academia Sinica (AS), Taipei, 11529, Taiwan\\
$^{2}$Institute of Communication Engineering (ICE), National Sun Yat-sen University (NSYSU), Kaohsiung, 80424, Taiwan\\
$^{3}$Department of Electronic Engineering, National Taipei University of Technology (NTUT), Taipei, 10608, Taiwan\\
$^{4}$Bradley Department of Electrical and Computer Engineering (ECE), Virginia Tech (VT),
Arlington, VA, 22203, USA\\
$^{5}$Artificial Intelligence \& Cyber Systems (AICS) Research Center, Lebanese American University (LAU), Lebanon\\
E-mails: d00942015@ntu.edu.tw, m113070007@student.nsysu.edu.tw, arthurchen0601@gmail.com, walids@vt.edu}
}

\maketitle

\begin{abstract}
In this paper, a novel analytical model for resource allocation is proposed for a device-to-device (D2D) assisted cellular network.
The proposed model can be applied to underlay and overlay D2D systems for sharing licensed bands and offloading cellular traffic.
The developed model also takes into account the problem of unlicensed band sharing with Wi-Fi systems.
In the proposed model, a global system state reflects the interaction among D2D, conventional cellular, and Wi-Fi packets.
Under the standard traffic model assumptions, a threshold-based flow control is proposed for guaranteeing the quality-of-service (QoS) of Wi-Fi.
The packet blockage probability is then derived.
Simulation results show the proposed scheme sacrifices conventional cellular performance slightly to improve overlay D2D performance significantly while maintaining the performance for Wi-Fi users.
Meanwhile, the proposed scheme has more flexible adjustments between D2D and Wi-Fi than the underlay scheme.
\end{abstract}


\section{Introduction}
The concept of device-to-device (D2D) communications, first proposed as part of LTE~\cite{Kato2015D2D, Release12}, could enable a plethora of wireless services such as vehicle-to-vehicle communications, while also improving spectrum efficiency. The gNodeB (gNB) base station can adopt D2D mode (one-hop link) and conventional cellular (CC) mode (two-hop link). D2D communication can be deployed either in the licensed band (in-band D2D) or in the unlicensed band (out-band D2D).~\cite{Liu2016D2D-U,Wu2016D2D-U,Zhang2017D2D-U,Kang2019D2D-U,Wu2019D2D-U,Saad2019D2D,Kato2020D2D-U,Han2021D2D-U}.

For in-band D2D communication, the allocation of the licensed band can be one of two types: \emph{overlay} in which the dedicated resources are orthogonally allocated to the D2D and CC modes in a way to eliminate the interference between D2D and CC links, or \emph{underlay} in which the D2D mode reuses the same resources as the CC mode thereby creating interference~\cite{Kato2015D2D,Han2021D2D-U} and~\cite{Iqbal2019D2D}. Since the underlay D2D and CC modes non-orthogonally share the same resources, the underlay D2D typically outperforms the overlay D2D in terms of spectral efficiency. However, sophisticated resource allocation schemes and interference management algorithms are needed for underlay D2D. These algorithms could incur excessive complexity and overhead at the gNB, particularly under dense deployment scenarios. To address this challenge, the interference can be reduced by gNB-assisted scheduling and the limitation of spectrum resources can be solved by out-band D2D communication. In particular, 3GPP 5th Generation (5G) Release 16~\cite{Release16} supports both license-assisted and stand-alone use of unlicensed spectrum in the 5 GHz and 6 GHz bands for 5G New Radio-Unlicensed (NR-U), respectively.

Out-band D2D communication can eliminate the interference between D2D and CC links and enhance the network capacity by extending D2D and CC communications on the unlicensed band such as the ISM band. However, out-band D2D requires sharing the unlicensed band with other wireless devices (e.g., Wi-Fi devices) which leads to uncontrollable interference between D2D and Wi-Fi links. Out-band D2D can be either \emph{controlled} (i.e., the resource allocation and interference avoidance are operated by the gNB in a centralized manner) or \emph{autonomous} (i.e., pairs of D2D connections are coordinated by users themselves)~\cite{Kato2015D2D,Lei2012D2D}. However, both controlled and autonomous D2D modes must be enabled by the gNB for purposes such as resource allocation or authentication. The protocol procedure for controlled and autonomous D2D was extensively detailed in~\cite{Lei2012D2D, D2Dcentralized}. For interference avoidance between out-band D2D and Wi-Fi, the listen-before-talk (LBT) mechanism was proposed by 3GPP LTE Release 13~\cite{Release13}. The LBT allows D2D users to operate the clear channel assessment (CCA) to check the presence of Wi-Fi packets on the target unlicensed band before transmission and prevent D2D packet collision with Wi-Fi packets. Meanwhile, the Wi-Fi system adopts a contention-based mechanism to avoid interference (i.e., CSMA/CA). The unlicensed band access adopted by the D2D technique is more flexible than the one adopted by the small cell~\cite{Wu2016D2D-U}.

\section{Related Works}

Most of the previous studies only focused on licensed band allocation between the CC and D2D modes~\cite{Lei2014D2D,Lei2016D2D,Liu2020}. However, the licensed band is limited and congested due to the spectrum reuse by inherent CC users, especially in the hotspot area. To improve the system capacity, D2D-U was proposed in~\cite{Zhang2017D2D-U,Wu2016D2D-U,Liu2016D2D-U}, and the LTE license-assisted access (LTE-LAA) technology~\cite{Chen2017LAA,Maule2018,Chou2019,Chou2020}. Prior works on D2D-U, like~\cite{Shang2017} and~\cite{Xing2020}, only discuss unlicensed band allocation, but the licensed band allocation between CC and D2D is not considered and can not reflect the practical scenario. In a D2D-U network, D2D users can share licensed bands with CC users and share unlicensed bands with Wi-Fi users. The average transmitter power and outage probability of the D2D-U link and the throughput of the arbitrary Wi-Fi user are theoretically analyzed in~\cite{Shang2017}. In~\cite{Xing2020}, the authors suggest a coexistence mechanism where Wi-Fi users accept some co-channel interference to enhance Wi-Fi security using D2D. Recently, several works~\cite{Liu2016D2D-U,Zhang2017D2D-U,Wu2019D2D-U,Han2021D2D-U} have investigated both licensed and unlicensed band allocation and performance analysis for D2D-U assisted cellular networks. These works focus on the interference management among D2D, CC, and Wi-Fi links with the goal of maximizing the throughput of D2D-U and CC networks while guaranteeing the quality of service (QoS) of Wi-Fi systems. The formulated optimization problems of these works are the mixed-integer non-linear programming (MINLP) problems, which are NP-hard problems. In~\cite{Wu2019D2D-U}, the authors maximize the total throughput of the network by optimizing the densities of the LTE, LTE-U, D2D, and D2D-U users.

Compared with the physical layer work~\cite{Wu2019D2D-U} that measures the throughput under an \emph{infinite backlog model} assumption, our work measures the system performance by the blocking probability under a \emph{standard traffic model} assumption. The work~\cite{Wu2019D2D-U} does not quantify Wi-Fi performance sacrifice for D2D-U and LTE-U enhancement. Inter-operation impacts CC and Wi-Fi performance, making inter-operation management crucial for easing data congestion. Furthermore, the work~\cite{Wu2019D2D-U} only considers the throughput performance of either uplink or downlink. However, this way is ideal from the medium access control (MAC) layer perspective. In contrast, we consider CC downlink and uplink simultaneously.
In~\cite{Lei2014D2D} and~\cite{Georgiadis2006}, the authors pointed out that resource allocation algorithms optimized under the infinite backlog model and considering only channel state information (CSI) may not guarantee QoS requirements for data traffic. They argued that these algorithms are not sufficient to handle packet QoS requirements under \emph{traffic model} assumptions. Further, the queue state information should also be considered in order to reflect the interaction in the real systems and the data packet traffic offloading. Markov queueing models are used in~\cite{Lei2014D2D} and~\cite{Lei2016D2D} for underlay and overlay D2D, assuming dynamic packet arrivals and departures with finite queues. However, they only consider licensed band sharing. The work in~\cite{Kang2019D2D-U} investigated the band allocation problem for underlay D2D using Markov analysis. However, this prior work used a random binary selection method, which can be inefficient because of a traffic re-steering scheme for diverting onto the licensed band when QoS failure of a session on the unlicensed band occurs. 

To the best of our knowledge, there are few mathematical models to characterize both licensed and unlicensed band allocation of D2D underlying/overlaying cellular networks under a \emph{traffic model} with flow admission control. The band allocation metric should depend on the band access tendency of the UEs, which in turn depends on the traffic arrival pattern. Thus, based on the observation of~\cite{Wu2019D2D-U} that ``\emph{users tend to access the unlicensed spectrum in a low traffic network, while new users prefer the licensed spectrum as the network traffic increases}'', we propose a threshold-based flow control with this principle in this work. Moreover, the proposed mathematical model based on the MAC layer can be applied to hybrid underlay and overlay D2D-U systems, which is an under-explored challenge in the existing literature~\cite{Lei2014D2D, Lei2016D2D} and~\cite{Kang2019D2D-U}.

The main contribution is to model and propose the threshold-based mechanism for licensed and unlicensed band allocation of underlay/overlay D2D packets under a \emph{traffic model}. In particular, we consider the gNB base station's data offloading operations for pairs of UEs in a centralized control approach. The decisions regarding packet traffic steering are made by selecting between different modes and bands: (a) CC mode (uplink and downlink) on the licensed band, (b) underlay/overlay D2D mode on the licensed band, and (c) D2D mode on the unlicensed band (D2D-U). For this heterogeneous network, we propose a link-layer analytical model to study the spectrum sharing issue for both underlay and overlay D2D. Then, in the proposed model, we model the interaction among D2D, CC, and Wi-Fi packets by using a global system state to reflect the practical traffic offloading. We show how much the performance of CC and Wi-Fi must be sacrificed for the improvement of D2D-U in the different schemes, and we propose a threshold-based flow control for D2D coexisting with Wi-Fi systems. 
Simulation results compare the performance of the proposed scheme with two baselines (overlay and underlay). 
It is shown that the proposed scheme outperforms the overlay scheme in D2D performance significantly by trading off CC performance slightly while maintaining Wi-Fi performance. In addition, although the proposed scheme is outperformed by the underlay scheme slightly, the underlay scheme sacrifices Wi-Fi performance significantly.

\section{Mode Selection and Band Allocation}
\begin{figure}[t]
\centering
{\includegraphics[width=0.4\textwidth]{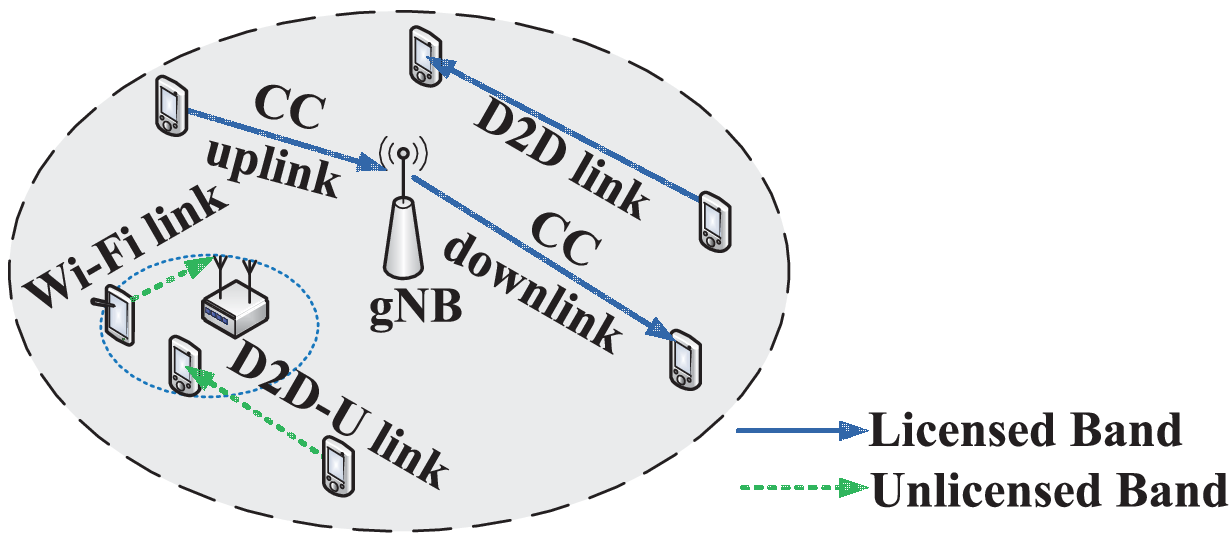}
\caption{Coverage area of a gNB.}\label{Region}}
\vspace{-0.15in}
\end{figure}

Consider a heterogeneous network, where the NR-U gNB and Wi-Fi access points serve their UEs, as shown in Fig.~\ref{Region}. When a new packet arrives in the gNB, the gNB selects either the CC or D2D mode for the transmission.
If the gNB selects the D2D mode for the packet, it selects either a licensed or unlicensed band and either underlay or overlay by performing threshold-based flow control.
If the channel capacity between the UEs is larger than the \emph{minimum} throughput requirement $R_{\textrm{min}}$, the gNB selects the D2D mode; otherwise, it chooses the CC mode.
That is, the gNB selects the D2D mode when the following condition meets 
\begin{align}
B_r\log_{2}\left(1+\gamma\right) \geq R_{\min},
\label{channel_capacity}
\end{align}
where $B_r$ is the residual bandwidth at the present time and $\gamma$ is the instantaneous signal-to-interference-plus-noise ratio (SINR) of the channel between UEs.
We assume that the packet requests within the coverage area of a gNB follow Poisson arrivals at rate $\Lambda$.
The probability of D2D mode selection will be $\rho$.
Therefore, the arrival rates of the mobile UEs' packet requests by using D2D and CC modes are $\Lambda_{A} = \rho \Lambda$ and $\Lambda_{B} = (1-\rho)\Lambda$, respectively.
The Wi-Fi packet is assumed to have  Poisson arrivals with rate $\Lambda_{C}$.

When the gNB selects the CC mode, the licensed band is allocated.
If the gNB selects the D2D mode, it decides either licensed or unlicensed bands and either underlay or overlay.
The gNB can assign the bit rate that satisfies the minimum QoS by allocating bandwidth to the UE.
Let $R_{\ell} = B_{\ell}\cdot \mathbb{E}[\log_{2}(1+\gamma_{\ell})]$ and $R_{u} = B_{u}\cdot \mathbb{E}[\log_{2}(1+\gamma_{u})]$ be, respectively, the residual bit rates for D2D packets, where $\gamma_{\ell}$ and $\gamma_{u}$ are the signal-to-noise ratio (SNR) of the licensed and unlicensed channel, $\mathbb{E}[ \cdot]$ is the expected operation, and $B_{\ell}$ and $B_{u}$ are the residual bandwidth of the licensed and unlicensed bands, respectively.
$\theta_{\ell}$ and $\theta_{u}$ are the preset thresholds for flow admission control in the licensed and unlicensed bands, respectively.
$R_{dd, T}$ is the total bit rate that can be allocated to the dedicated licensed channel for overlaying D2D packets by the gNB.
$R_{dw, T}$ and $R_{up, T}$ represent the total bit rate that can be allocated to the uplink and downlink licensed channel for CC packets by the gNB.
Since $R_{up, T}$ can be shared with underlay D2D packets, the threshold is set as $\theta_{\ell} = R_{up, T} - R_{dd, T}$.
$R_{u, T}$ is the total bit rate that can be allocated to Wi-Fi packets and can be shared with D2D-U packets.
Since $R_{u, r}$ is reserved for Wi-Fi packets only, the threshold is set as $\theta_{u} = R_{u, T} - R_{u, r}$.
We consider five cases:
\textbf{Case 1:} $R_{\ell} \geq \theta_{\ell}$, $R_{u} \geq \theta_{u}$. If both the residual licensed and unlicensed bands can sufficiently support the required transmission rate of the incoming packet, then the D2D packet will be allocated to the unlicensed band.
\textbf{Case 2:} $R_{\ell} \geq \theta_{\ell}$, $R_{u} < \theta_{u}$. If the unlicensed band is congested by Wi-Fi and other D2D packets, then the new D2D arrival packet accesses the dedicated licensed band, \emph{overlaying} the cellular network.
\textbf{Case 3:} $R_{\ell} < \theta_{\ell}$, $R_{u} \geq \theta_{u}$. If the licensed band is almost occupied by other D2D and CC packets, then the new D2D arrival packet uses the unlicensed band.
\textbf{Case 4:} $R_{\ell} < \theta_{\ell}$, $R_{u} < \theta_{u}$, and $R_{\ell} > 0$. If both the residual licensed and unlicensed bands are not sufficient for the required transmission rate of the incoming packet, then the D2D packet underlays the cellular network and competes with the CC packets in the residual licensed band.
\textbf{Case 5:} $R_{\ell} \simeq 0$, $R_{u} < \theta_{u}$. If the licensed and unlicensed bands are fully loaded, then the D2D packet is blocked.

For offloading the data packets on licensed and unlicensed bands simultaneously under the \emph{traffic model} assumption, we use Cases 1 and 2 to allocate bands in the low-traffic situation and adopt Cases 3 and 4 to allocate bands in the high-traffic situation based on the suggestion of~\cite{Wu2019D2D-U}. The threshold mechanism can reduce D2D systems to underlay or overlay schemes. In Case 4, if we let $\theta_{\ell} = 0$ and $R_{dd, T} = R_{up, T}$, then the D2D system is reduced to an underlay scheme. Meanwhile, if we let $\theta_{\ell} > 0$, then the D2D system is reduced to an overlay scheme. On the other hand, the gNB needs to reserve part of the unlicensed band by the preset threshold $\theta_{u}$ in Cases 2, 4, and 5 to guarantee the QoS of Wi-Fi on the unlicensed band. Therefore, we propose an analytic model to evaluate the packet blocking probability while the threshold mechanism is enabled. It can be applied to the hybrid underlay and overlay D2D system, which is seldom discussed in previous link-layer studies.

\section{Analytical Model}

\begin{figure}[t]
\centering
{\includegraphics[width=0.5\textwidth]{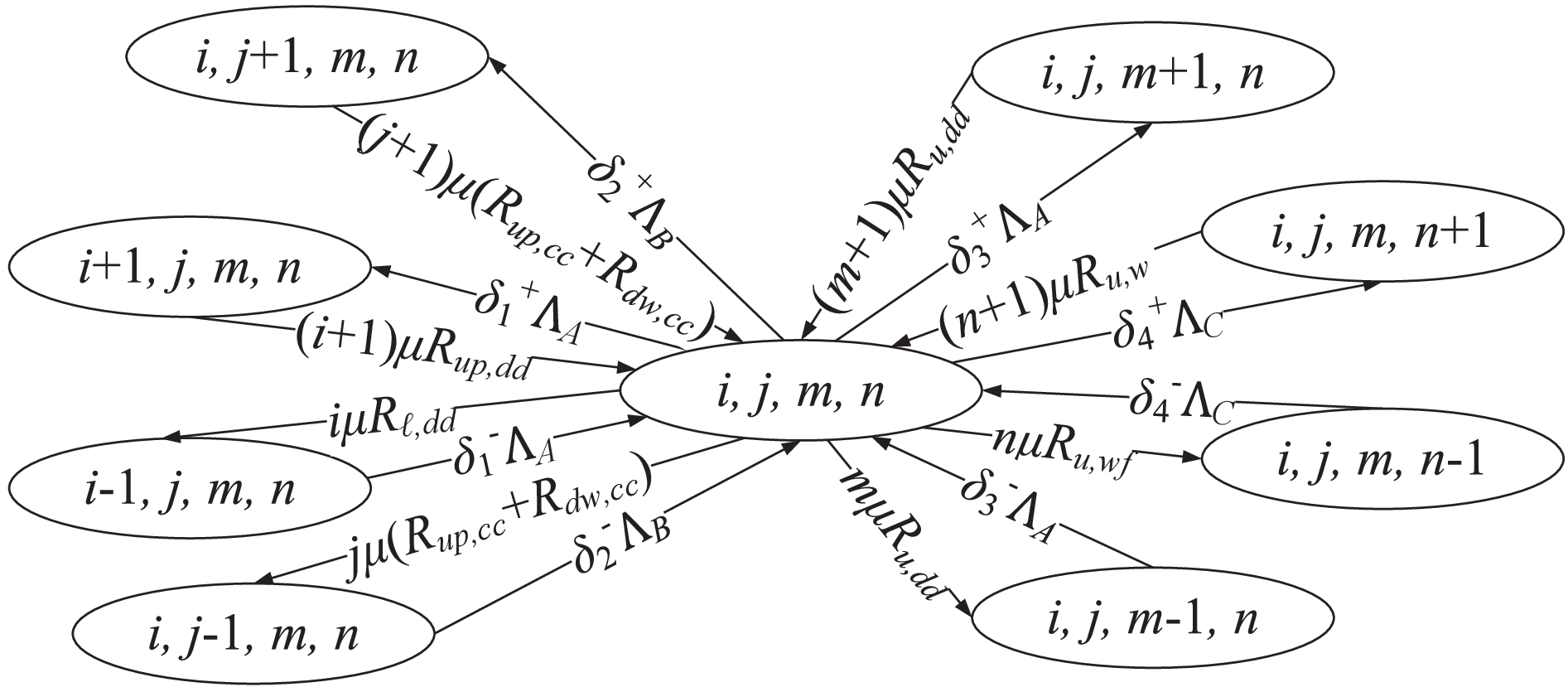}
\caption{The state transition diagram for the underlay/overlay D2D network.}
\label{States}}		
\vspace{-0.15in}
\end{figure}
We now propose an analytical model to investigate the performance of the unlicensed and licensed band allocation for D2D underlaying or overlaying cellular networks.
We define $X$ as a random variable that represents the packet size of the requests from the coverage area of a gNB. $X$ has an exponential distribution with mean $1/\mu$ and density function $f(x) = \mu e^{- \mu x}$.
When the packet is delivered with an average bit rate $R$, the average transmission time required will be $X/R$, which also has an exponential distribution with mean $1/R\mu$ and density function $f(x) = R\mu e^{- R\mu x}$.
Based on queueing theory, the steady state of a queueing system is the state where the probability of the number of packets in the system is independent of time.
To simplify the analysis, we consider only single-level average bit rates that satisfy minimum QoS in the same protocol service.
The proposed model can be applied to multiple-level bit rate services, as explained in~\cite{PLin2004}.

The behaviors of D2D, CC, and Wi-Fi networks can be modeled by a 4-D Markov process.
A state $(i, j, m, n)$ indicates that within the coverage area of a gNB,
$i$ packets are being delivered with an average bit rate $R_{\ell, dd}$ transmissions by using D2D on the uplink licensed channel.
$j$ packets are being delivered with average bit rates $R_{up, cc}$ and $R_{dw, cc}$ by using CC on the uplink and downlink licensed channels, respectively.
$m$ packets are being delivered with an average bit rate $R_{u, dd}$ transmissions by using D2D on an unlicensed channel.
$n$ packets are being delivered with an average bit rate $R_{u, wf}$ transmissions by using Wi-Fi on an unlicensed channel. We assume that the Wi-Fi data packet is one that has won the competition over the media contention protocol CSMA/CA in the Wi-Fi network.
Fig.~\ref{States} illustrates the transition diagram for this Markov process.

The state space $\mathcal{S}$ for the Markov process is expressed as
\begin{align}
&\mathcal{S} = \Bigg\{ (i, j, m, n) \; | \; 0 \leq i R_{\ell, dd} \leq R_{dd, T}, 0 \leq j R_{dw, cc} \leq R_{dw, T}, \notag\\
&0 \leq i R_{\ell, dd} + j R_{up, cc} \leq R_{up, T},
0 \leq m R_{u, dd} + n R_{u, wf} \leq R_{u , T}, \notag\\
&0 \leq i \leq \left\lfloor \frac{R_{up, T}}{R_{\ell, dd}} \right\rfloor + \left\lfloor \frac{R_{dd, T}}{R_{\ell, dd}} \right\rfloor, 0 \leq m \leq \left\lfloor \frac{R_{u, T}-R_{u, r}}{R_{u, dd}} \right\rfloor ,\notag\\
&0 \leq j \leq \left\lfloor \min\left( \frac{R_{up, T}}{R_{up, cc}}, \frac{R_{dw, T}}{R_{dw, cc}} \right)\right\rfloor,  0 \leq n \leq \left\lfloor \frac{R_{u, T}}{R_{u, wf}} \right\rfloor\Bigg\}.
\label{State_Space}
\end{align}
Let $\pi_{i, j, m, n}$ be the steady state probability for state $(i, j, m, n)$, where $\pi_{w,x,y,z} = 0$ if state $(i, j, m, n) \notin \mathcal{S}$.
For all legal states $(i, j, m, n) \in \mathcal{S}$, $\sum_{(i, j, m, n) \in \mathcal{S}} \pi_{w,x,y,z} = 1$.
The balance equations for this Markov process will be:
\begin{align}
&\pi_{i, j, m, n} = \Big\{ (i+1)\mu R_{\ell, dd}\pi_{i+1, j, m, n} \notag\\
& + (j+1)\frac{\mu}{2} (R_{up, cc} + R_{dw, cc})\pi_{i, j+1, m, n} \notag\\
& + (m + 1)\mu R_{u, dd}  \pi_{i, j, m+1, n} + (n + 1)\mu R_{u , wf}\pi_{i, j, m, n+1}\notag\\
& + \delta_{1}^{-}\pi_{i-1, j, m, n} + \delta_{2}^{-}\pi_{i, j-1, m, n} + \delta_{3}^{-}\pi_{i, j, m-1, n}\notag\\
& + \delta_{4}^{-}\pi_{i, j, m, n-1}\Big\}\times \Big[\delta_{1}^{+}\Lambda_{A} + \delta_{2}^{+}\Lambda_{B} + \delta_{3}^{+}\Lambda_{A} + \delta_{4}^{+}\Lambda_{C} +\notag\\
& i\mu R_{\ell, dd} + j\frac{\mu}{2}(R_{up, cc} + R_{dw, cc}) + m\mu R_{u, dd} + n\mu R_{u , wf}\Big]^{-1}.
\label{Balance Equation}
\end{align}
For state $(i, j, m, n)$, we consider the following transitions between $(i+1, j, m, n)$, $(i, j+1, m, n)$, $(i, j, m+1, n)$, $(i, j, m, n+1)$ as shown in Fig.~\ref{States}.
\begin{itemize}
\item[1)] When the unlicensed band has heavy traffic (i.e., $m R_{u, dd} + n R_{u, wf} > \theta_{u}-R_{u, dd}$), and a D2D packet request arrives at state $(i, j, m, n)$ where the residual bit rate of the licensed band is enough to support D2D transmission (i.e., $i R_{\ell, dd} + j R_{up, cc} \leq R_{up, T} - R_{\ell, dd}$), then the request is allocated one D2D licensed channel.  Therefore, the process moves from $(i, j, m, n)$ to state $(i+1, j, m, n)$ with rate $\delta_{1}^{+}\Lambda_{A}$, where
\begin{align}
\delta_{1}^{+} &= \left\{
\begin{array}{ll}
1, &0\leq i R_{\ell, dd} + j R_{up, cc} \leq R_{up, T} - R_{\ell, dd},\\
&m R_{u, dd} + n R_{u, wf} > \theta_{u}-R_{u, dd}, \\
&\textrm{and} \;\; (i, j, m, n) \in \mathcal{S}\\
0, &\textrm{otherwise}.
\end{array}\right.\notag
\end{align} 


\item[2)] When a D2D packet transmission on the licensed channel for the request is completed at state $(i+1, j, m, n)$, one licensed channel with D2D is released. The process moves from state $(i+1, j, m, n)$ to $(i, j, m, n)$ with rate $(i+1)\mu R_{up,dd}$.

\item[3)] For a CC packet request at state $(i, j, m, n)$, when the residual bit rate of uplink and downlink licensed bands is sufficient to support CC transmission (i.e., $j R_{up,cc}\leq R_{up,T}-R_{up,cc}$ and $j R_{dw,cc}\leq R_{dw,T}-R_{dw,cc}$), then the request is allocated one CC uplink licensed channel and one CC downlink licensed channel. Therefore, the process moves from $(i, j, m, n)$ state to $(i+1, j, m, n)$ with rate $\delta_{2}^{+}\Lambda_{B}$, where
\begin{align}
\delta_{2}^{+} &= \left\{
\begin{array}{ll}
1, &0\leq i R_{\ell, dd} + j R_{up,cc}\leq R_{up,T}-R_{up,cc},\\
&0\leq j R_{dw,cc}\leq R_{dw,T}-R_{dw,cc},\\
&\textrm{and} \;\; (i, j, m, n) \in \mathcal{S}\\
0, &\textrm{otherwise}.
\end{array}\right.\notag
\end{align}
\item[4)] When a CC packet transmission for the request completes at state $(i, j+1, m, n)$, one CC uplink licensed channel and one CC downlink licensed channel are released. The process moves from state $(i, j+1, m, n)$ to state $(i, j, m, n)$ with rate $(j+1)\mu (R_{up,cc}+R_{dw,cc})$.

\item[5)] When the unlicensed band has light traffic, and a D2D packet request arrives at state $(i, j, m, n)$ where the residual bit rate of the unlicensed band is sufficient to support D2D transmission (i.e., $m R_{u, dd} + n R_{u, wf}\leq \theta_{u}-R_{u, dd}$), then the request is allocated one D2D unlicensed channel. Therefore, the process moves from state $(i, j, m, n)$ state to $(i, j, m+1, n)$ with rate $\delta_{3}^{+}\Lambda_{A}$, where
\begin{align}
\delta_{3}^{+} &= \left\{
\begin{array}{ll}
1, &0\leq m R_{u, dd} + n R_{u, wf}\leq \theta_{u}-R_{u, dd},\\
& \textrm{and} \;\; (i, j, m, n) \in \mathcal{S}\\
0, &\textrm{otherwise}.
\end{array}\right.\notag
\end{align}
\item[6)] When a D2D packet transmission on the unlicensed channel for the request completes at state $(i, j, m+1, n)$, one unlicensed channel with D2D is released. The process moves from state $(i, j, m+1, n)$ to $(i, j, m, n)$ with rate $(m+1)\mu R_{u, dd}$.

\item[7)] For a Wi-Fi packet request at state $(i, j, m, n)$, if the residual bit rate of unlicensed band is enough to support Wi-Fi transmission (i.e., $m R_{u, dd} + n R_{u, wf}\leq R_{u, T}-R_{u, wf}$), then the request is allocated one Wi-Fi unlicensed channel. Therefore, the process moves from $(i, j, m, n)$ state to $(i, j, m, n+1)$ with rate $\delta_{4}^{+}\Lambda_{C}$, where 
\begin{align}
\delta_{4}^{+} &= \left\{
\begin{array}{ll}
1, &0\leq m R_{u, dd} + n R_{u, wf}\leq R_{u,T}-R_{u, wf},\\
&\textrm{and} \;\; (i, j, m, n) \in \mathcal{S}\\
0, &\textrm{otherwise}.
\end{array}\right.\notag
\end{align}
\item[8)] When a Wi-Fi packet transmission on the unlicensed channel for the request completes at state $(i, j, m, n+1)$, one unlicensed channel with a Wi-Fi bit rate is released. The process moves from state $(i, j, m, n+1)$ to $(i, j, m, n)$ with rate $(n+1)\mu R_{u, wf}$.
\end{itemize}

The transitions between $(i, j, m, n)$ and $(i-1, j, m, n)$, $(i, j-1, m, n)$, $(i, j, m-1, n)$, $(i, j, m, n-1)$ are similar to those between $(i, j, m, n)$ and $(i+1, j, m, n)$, $(i, j+1, m, n)$, $(i, j, m+1, n)$, $(i, j, m, n+1)$, as shown in Fig.~\ref{States}. We can now obtain the transition conditions $\delta_{1}^{-}, \delta_{2}^{-}, \ldots , \delta_{4}^{-}$ as follows:
\begin{align}
\delta_{1}^{-} &= \left\{
\begin{array}{ll}
1, &0\leq (i-1) R_{\ell, dd} + j R_{up, cc} \leq R_{up, T} - R_{\ell, dd},\\
&m R_{u, dd} + n R_{u, wf} > \theta_{u}-R_{u, dd},\\
&\textrm{and} \;\; (i, j, m, n) \in \mathcal{S}\\
0, &\textrm{otherwise}.
\end{array}\right.\notag
\end{align}

\begin{align}
\delta_{2}^{-} &= \left\{
\begin{array}{ll}
1, &0\leq i R_{\ell, dd} + (j-1) R_{up,cc}\leq R_{up,T}-R_{up,cc},\\
&0 \leq (j-1) R_{dw,cc}\leq R_{dw,T}-R_{dw,cc},\\
&\textrm{and} \;\; (i, j, m, n) \in \mathcal{S}\\
0, &\textrm{otherwise}.
\end{array}\right.\notag\\
\delta_{3}^{-} &= \left\{
\begin{array}{ll}
1, &0\leq (m-1) R_{u, dd} + n R_{u, wf}\leq \theta_{u}-R_{u, dd},\\
&\textrm{and} \;\; (i, j, m, n) \in \mathcal{S}\\
0, &\textrm{otherwise}.
\end{array}\right.\notag\\
\delta_{4}^{-} &= \left\{
\begin{array}{ll}
1, &0\leq m R_{u, dd} + (n-1) R_{u, wf}\leq R_{u,T}-R_{u, wf},\\
&\textrm{and} \;\; (i, j, m, n) \in \mathcal{S}\\
0, &\textrm{otherwise}.
\end{array}\right.\notag
\end{align}


A packet request is blocked if the residual bit rate is not enough to support it and thus, we can define the D2D, CC, and Wi-Fi packets blocking probabilities as follows:
\begin{align}
\label{block_prob_D2D}
P_{b, dd} &= \sum_{\begin{array}{c}
(i, j, m, n) \in \mathcal{S},\\
i R_{\ell, dd} + j R_{up, cc} > R_{up, T}-R_{\ell, dd},\\
m R_{u, dd} + n R_{u, wf} > \theta_{u}-R_{u, dd}
\end{array}}\pi_{i, j, m, n},\\
\label{block_prob_CC}
P_{b, cc} &= \sum_{\begin{array}{c}
(i, j, m, n) \in \mathcal{S},\\
\left(i R_{\ell, dd} + j R_{up,cc}> R_{up,T}-R_{up,cc}\right)\\
\cup\left(j R_{dw,cc}> R_{dw,T}-R_{dw,cc}\right)
\end{array}}\pi_{i, j, m, n},
\end{align}
and
\begin{align}
\label{block_prob_WF}
P_{b, wf} &= \sum_{\begin{array}{c}
(i, j, m, n) \in \mathcal{S},\\
m R_{u, dd} + n R_{u, wf} > R_{u, T}-R_{u, wf}
\end{array}}\pi_{i, j, m, n}.
\end{align}
For computing the steady-state probabilities, we use the iterative approach shown as \textbf{Algorithm 1} that has been widely used and validated by experiments~\cite{Chou2019,Chou2020} and~\cite{PLin2004}.
The proof in~\cite{Fu2013,ACPang2001} had shown that the iterative algorithm for steady-state probability converges to a unique sub-optimal solution.

\begin{algorithm}
\caption{Iterative algorithm}\label{alg:three}
\SetKwInput{KwData}{Input}
\SetKwInput{KwResult}{Output}
\KwData{$t = 1$ where $t$ indexes the iteration number.\\ The D2D mode selection probability $\rho$. The D2D, CC and Wi-Fi packets arrival rates $\Lambda_{A}$, $\Lambda_{B}$ and $\Lambda_{C}$. The converge threshold $\alpha = 10^{-6}$. The mean and total bit rates of D2D, CC and Wi-Fi $R_{\ell, dd}$, $R_{up, cc}$, $R_{dw, cc}$, $R_{u, dd}$, $R_{u, wf}$, $R_{dd, T}$, $R_{dw, T}$, $R_{up, T}$ and $R_{u, T}$.}
\KwResult{The D2D, CC and Wi-Fi packet blockage \\probability $P_{b, d2d}$, $P_{b, cc}$, and $P_{b, wf}$.}
Select initial values for all $\pi_{i, j, m, n}(t)$ by~(\ref{State_Space}) and~(\ref{Balance Equation})\;
 \While{$t \neq 0$}{
  Compute $\pi_{i, j, m, n}(t)$ by~(\ref{Balance Equation})\;
  Let $\pi_{i, j, m, n}(t+1) = \pi_{i, j, m, n}(t)$\;
  Compute $G^{-1}\pi_{i, j, m, n}(t+1)$, where the normalized factor $G=\sum_{(i, j, m, n) \in \mathcal{S}} \pi_{i, j, m, n}(t+1)$ is used to ensure that $\sum_{(i, j, m, n) \in \mathcal{S}} G^{-1}\pi_{i, j, m, n}(t+1) = 1$\;
  \eIf{$|\pi_{i, j, m, n}(t+1) - \pi_{i, j, m, n}(t)| \leq \alpha$}{
   Compute $P_{b, dd}$, $P_{b, cc}$, and $P_{b, wf}$ by~(\ref{block_prob_D2D})-(\ref{block_prob_WF})\; Let $t = 0$\;
   }{
   Let $t = t + 1$\;
  }
 }
\end{algorithm}

\section{Performance Evaluation}
\begin{figure}[h]
\centering
{\includegraphics[width=0.55\textwidth]{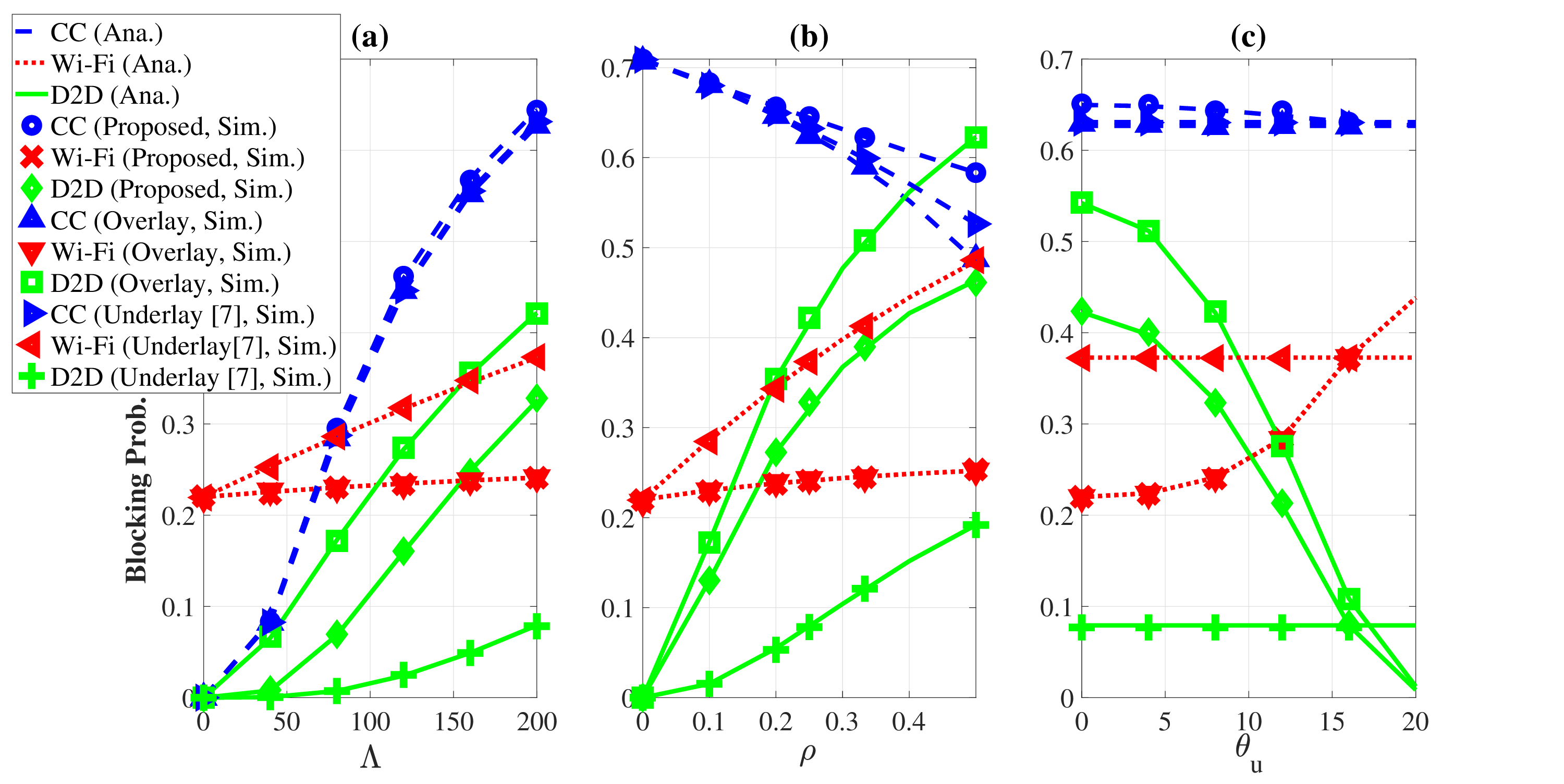}
\caption{The effects of (a) $\Lambda$ (units: $r\mu$), (b) $\rho$, and (c) $\theta_{u}$ (units: $r$).}
\label{Results}}
\vspace{-0.15in}
\end{figure}

In this section, we validate the proposed analytical model via simulation experiments.
The simulation experiments follow the discrete event-driven simulation approach in~\cite{Chou2019,Chou2020,PLin2004}.
Based on the file transfer protocol (FTP) standard traffic model in~\cite{Release12,Release13}, we study the packet blocking probability of the D2D, CC, and Wi-Fi.
In our study, the mean of packet inter-arrival times $1/\Lambda$ is normalized by the mean of the D2D transmission times on the licensed band for a packet $1/r\mu$ (i.e., if the average transmission time for a D2D packet is $1/r\mu = 0.4$s, then $\Lambda = 20r\mu$ means that the average packet inter-arrival time is $1/\Lambda = 1/20r\mu = 0.02$s). Here, we neglect the small CCA period ($20\mu$s)~\cite{Release13}. In~\cite{Release12}, the bit rate of the codec is $r = 12.2$kb/s.
We set $r = R_{\ell, dd}$ to simplify the discussion. Fig.~\ref{Results} plots the blocking probabilities as functions of $\Lambda$, $\rho$, and $\theta_{u}$, where $\Lambda_{C} = 100r\mu$, $R_{dd,T} = 2r$, $R_{up,T} = 4r + R_{dd,T}$, $\theta_{\ell} = 4r$, $ R_{dw,T} = 4r$, $R_{u,T} = 8r$, $R_{u,dd} = R_{u,wf} = 2r$, and $R_{\ell,up} = R_{\ell,dw} = r$. When $\Lambda$, $\rho$, or $\theta_{u}$ are not variables on the horizontal axis, $\Lambda = 200r\mu$, $\rho = 1/4$, and $\theta_{u} = 4r$.
In Fig.~\ref{Results}(a), when $\Lambda$ increases, $P_{b, dd}$ and $P_{b, cc}$ increase dramatically, but $P_{b, wf}$ increases slightly, which suggests that the licensed band is preferred when the traffic is heavy. This scenario reflects that the new packet prefers the licensed band as the network traffic increases~\cite{Wu2019D2D-U}. This is because the proposed threshold-based flow control reserves the residual unlicensed band by the preset threshold $\theta_{u}$ for guaranteeing the QoS of Wi-Fi on the unlicensed band.
In Fig.~\ref{Results}(b), when $\rho$ (the selected probability of D2D mode) increases, $P_{b, dd}$ increases, but $P_{b, cc}$ decreases, which indicates the traffic is offloaded flexibly by D2D. In Fig.~\ref{Results}(c), when $\theta_{u}$ (the unlicensed bandwidth that can be used by D2D) increases, $P_{b, wf}$ increases, but $P_{b, dd}$ decreases, which indicates the impact on Wi-Fi is adjusted flexibly by using flow control and $\theta_{u}$ has a slight impact on CC performance. When $\theta_{u} = 12r$, the performance of D2D and Wi-Fi reaches a balanced relationship. The blocking probability of D2D can be adjusted from 0 to 0.6 by threshold $\theta_{u}$, while the blocking probability of Wi-Fi is adjusted almost twice as much as the original.
We compare the proposed scheme with two baselines: \textbf{(1) Overlay:} D2D and CC use only the dedicated resources in the licensed band, respectively. D2D uses flow control in the unlicensed band as well. In other words, we remove Case 4 from the proposed scheme. \textbf{(2) Underlay:} D2D and CC compete for resources in the licensed band, and D2D does not adopt flow control in the unlicensed band, as described in~\cite{Wu2019D2D-U}. As shown in Fig.~\ref{Results}(a) and Fig.~\ref{Results}(b), the proposed scheme sacrifices CC performances ($2.7\%$ and $19.7\%$) for improving overlay D2D performances ($22.1\%$ and $25.9\%$), while $\Lambda = 200r\mu$ and $\rho = 0.5$, respectively. Meanwhile, the underlay D2D sacrifices CC ($0.6\%$ and $8\%$) and Wi-Fi performances ($54.7\%$ and $92.9\%$) to outperform overlay D2D ($81.5\%$ and $69.1\%$), respectively. We find that the underlay D2D outperforms the proposed scheme ($76.2\%$ and $58.4\%$) by sacrificing the Wi-Fi performances ($54.8\%$ and $92.9\%$). Meanwhile, the proposed scheme has no sacrificing on Wi-Fi performance compared to underlay D2D. In Fig.~\ref{Results}(c), the proposed scheme sacrifices CC performance ($3.2\%$) to improve overlay D2D ($22.2\%$), while $\theta_{u} = 0r$. Meanwhile, the underlay D2D sacrifices the Wi-Fi performance (69.9\%) to outperform the proposed and overlay D2D schemes (81.8\% and 85.8\%).
However, the underlay scheme of~\cite{Wu2019D2D-U} cannot adjust the data traffic. It is worth mentioning that the proposed scheme easily achieves the performance balance between D2D and Wi-Fi, while $\theta_{u} = 10r$.

\section{Conclusion}
In this paper, we have proposed an analytical model for resource allocation in D2D networks with simulation validation.
This model can be applied to both underlay and overlay D2D for sharing licensed bands with CC packets.
To guarantee the QoS of Wi-Fi users, we have developed a threshold-based flow control solution for D2D sharing unlicensed bands.
Performance results indicate that the setting of D2D mode and flow control increases the flexibility of offloading cellular traffic while impacting Wi-Fi users slightly.
Based on the standard traffic model, the proposed solution provides guidelines to network operators on how to set the available resources for a D2D system.

\bibliographystyle{IEEEtran}

\end{document}